# Inherent properties of binary tetrahedral semiconductors

**A. S. Verma**[*1], B. K. Sarkar[2] and V. K. Jindal[1]

1 Department of Physics, Panjab University, Chandigarh 160014, India
2 Department of Physics, VIT University, Vallore 632014, India

**Abstract:** A new approach utilising the concept of ionic charge theory has been used to explain the inherent properties such as lattice thermal conductivity and bulk modulus of $A^{III}B^{V}$ and $A^{II}B^{VI}$ semiconductors. The lattice thermal conductivity (K) of these semiconductors exhibit a linear relationship when plotted on a log–log scale against the nearest neighbour distance *d* (°A), but fall on two straight lines according to the product of the ionic charge of the compounds. On the basis of this result a simple lattice thermal conductivity–bulk modulus relationship is proposed and used to estimate the bulk modulus of these semiconductors. A fairly good agreement has been found between the experimantal and calculated values of these parameters for zinc blende structured solids.

[**Keywords:** A. Semiconductors; D. Thermal Properties]

## 1 Introduction

Most of the physical world around us and a large part of modern technology are based on solid materials. The extensive research devoted to the physics and chemistry of solids during the last quarter of a century has led to great advances in understanding of the properties of solids in general. So it is interesting to study the behaviour and various properties of different solids. In recent years there has been considerable interest in theoretical and experimental studies of $A^{N}B^{8-N}$ type crystals with zinc blende structure. It is attributed to their high symmetry and simplicity of their ionic bonding [1]. Almost all the $A^{II}B^{VI}$ and $A^{III}B^{V}$ compounds crystallize either in the zinc blende or wurtzite structures. The common and dominant feature of these structures is the tetrahedral bonding to four atoms of the other elements. In zinc blende these tetrahedral are arranged in a cubic type structure whilst they are in a hexagonal type structure. Indeed, the centres of similar tetrahedral are arranged in a face-centred cubic (*fcc*) array in the former and a hexagonal closed-packed (*hcp*) array in the latter

- Corresponding author: e-mail: ajay_phy@rediffmail.com, Mob: +91 9412884655,

[2]. The particular omnitriangulated nature in atomic structure gives these materials unique physical properties. During the last few years, a number of theoretical calculations based on empirical relations have become an essential part of material research. Because ab initio calculations are complex and required significant effort, more empirical calculations [3, 4], have been developed to compute properties of zinc blende solids. The empirical relations have become widely recognized as the method of choice for computational solid-state studies. In modern high-speed computer techniques, it allows researchers to investigate many structural and physical properties of materials only by computation or simulation instead of traditional experiments. Empirical concepts such as valence, empirical radii, electronegativity, ionicity and plasmon energy are then useful [5, 6]. These concepts are directly associated with the character of the chemical bond and thus provide means for explaining and classifying many basic properties of molecules and solids.

The valence electrons refer to the electrons that take part in chemical bonding. Recently, Verma and co-authors [7-12] have calculated the electronic, mechanical and optical properties of binary and ternary compounds with the help of ionic charge theory of solids. This is due to the fact that the ionic charge depends on the number of valence electrons, which changes when a metal forms a compound. Therefore we thought it would be of interest to give an alternative explanation for some inherent properties such as lattice thermal conductivity and bulk modulus of zinc blende solids.

## 2 Theory, results and discussion

Lattice thermal conductivity and bulk modulus need careful investigation as they are correlated with cohesivity, nature of covalency, bond ionicity and electronic behaviour of the constituent element forming the compound [13,14]. Spitzer [15] and Ioffe [16], experimentally estimated the lattice thermal conductivity (K) of binary and ternary crystals. Different theoretical models, based on bond ionicity, and melting temperature, have been proposed by several other researchers [13-16]. The bond ionicity theory of solids has been used by Garbato et al [13] for the calculation of K of the tetrahedral semiconductors and lattice thermal conductivity (K) of these semiconductors may be expressed as,

$$K = K^*_{cov} (1-f_i/F_i) (T^*/T) \quad (1)$$

where $T^*$ is the melting temperature, $f_i$ is the crystal ionicity, $F_i$ is the critical ionicity.

In the previous work, [7-12], we have proposed simple expressions for the electronic, optical and mechanical properties such as heteropolar energy gaps ($E_c$), average energy gaps ($E_g$), crystal ionicity ($f_i$), dielectric constant ($\varepsilon_\infty$), electronic susceptibility ($\chi$), cohesive energy ($E_{coh}$), bulk modulus

(B) and microhardness (H) of rocksalt, zinc blende and complex structured solids in terms of the product of ionic charges of cation and anion by the following relations,

$$f_i = 0.87891 \times d^{0.24} / (Z_1 Z_2)^{0.4}, \quad (2)$$

The crystal ionicity ($f_i$) depends on the product of ionic charges [9]. Thus, there must be a correlation between K and product of ionic charges. The lattice thermal conductivity of zinc blende solids exhibit a linear relationship when plotted against nearest-neighbour distance, but fall on two straight lines according to the ionic charge product of the compounds, this is presented in figure 1. We observe that in the plot of lattice thermal conductivity and nearest neighbour distance; the $A^{II}B^{VI}$ semiconductors lie on line nearly parallel to the line for the $A^{III}B^{V}$ semiconductors. From the figure 1 it is quite obvious that the lattice thermal conductivity trends in these compounds decreases with increases nearest neighbour distance and fall on different straight lines according to the ionic charge product of the compounds. We are of the view that lattice thermal conductivity (K in W/K cm) of these semiconductors may be evaluated using their ionic charge by the following relation,

$$K = S\,(Z_1 Z_2)^{V} / d^{5} \quad (3)$$

Where $Z_1$ and $Z_2$ are the ionic charge of the A and B respectively and d is the nearest neighbour distance in Å. S and V are constants, which depends upon crystal structure, they have values of 2 and 1.5 respectively for zinc blende structured solids.

The bulk modulus defines its resistance to volume change when compressed. Both experimental and theoretical results suggest that the bulk modulus is a critical single material property to indicate hardness. Recently, Al-Douri et al. [1] have studied the bulk modulus of IV, III-V and II-VI semiconductors and proposed an empirical relation for bulk modulus (B in Gpa) in terms of transition pressure ($P_t$). According to them bulk modulus of these semiconductors may be expressed as,

$$B = [99 - (\lambda + 79)]\,(10P_t)^{1/3} \quad (4)$$

Where $P_t$ is the transition pressure in GPa from ZB to β-Sn and λ is a parameter appropriate for the group-IV (λ = 1), III-V (λ = 5) and II-VI (λ = 8) semiconductors. On the basis of lattice thermal conductivity result a simple lattice thermal conductivity—bulk modulus relationship is proposed and used to estimate the bulk modulus (B in Gpa) of these semiconductors. Therefore, we have plotted a graph between experimental bulk modulus and lattice thermal conductivity values, which is shown in figure 2 for the above series of compounds. From the figure 2 it is quite obvious that the $A^{II}B^{VI}$ and $A^{III}B^{V}$ semiconductor lies on two straight lines according to the ionic charge product of the compounds. Thus bulk modulus of these compounds may be evaluated by the following relation,

$$B = C\,K^{0.75} \qquad (5)$$

Where C is a constant, the numerical value of C is 110 and 235 for $A^{III}B^{V}$ and $A^{II}B^{VI}$ respectively. A detailed discussion of this parameter for these materials has been given elsewhere [13-19] and will not be presented here. Using equations (3) and (5) lattice thermal conductivity and bulk modulus for $A^{II}B^{VI}$ and $A^{III}B^{V}$ semiconductors have been calculated. The results are presented in table 1. The calculated values are in good agreement with the experimental values reported by earlier researchers [13,17-19].

## 3 Conclusions

In the proposed models, calculations are simple, fast and more accurate; in regards of the applications point of view it can be highly dependent. The only information needed for calculating lattice thermal conductivity and bulk modulus by proposed method is the nearest neighbour distance and ionic charge and evaluated values are in better agreement with experiment data as compared to empirical relations proposed by previous researchers [13-19]. We come to the conclusion that ionic charge of any compound is a key parameter for calculating the electronic, optical and mechanical properties. It is natural to say that this model can easily be extended to complex crystals for which the work is in progress and will be appearing in forthcoming paper.

**Acknowledgements**: One of the authors (Dr. A. S. Verma, PH/08/0049) is thankful to the University Grant Commission New Delhi, India for supporting this research under the scheme of U.G.C. Dr. D.S. Kothari Post Doctoral Fellowship.

## References

[1] A. Mujica, A. Rubio, A. Munoz and R. J. Needs, Rev. Mod. Phys. **75** (2003) 863.

[2] Mukesh Jain, Diluted magnetic semiconductors (Singapore: World Scientific), (1991).

[3] H. M. Tutuncu, S. Bagci, G. P. Srivastava, A. T. Albudak and G. Ugur, Phys. Rev. B **71** (2005) 195309.

[4] R.R. Reddy, K.R. Gopal, K. Narasimhulu, L.S.S. Reddy, K.R. Kumar, G. Balakrishnaiah and M. R. Kumar, J. Alloys Compd. **473** (2009) 28.

[5] L. Pauling, The Nature of the Chemical Bond, 3rd, ed. (Cornell University Press, Ithaca, 1960).

[6] V. Kumar, and B. S. R. Sastry, J. Phys. Chem. Solids **66** (2005) 99.

[7] A. S. Verma, Physica Scripta **79** (2009) 045703.

[8] A. S. Verma and S. R. Bhardwaj, J. Phys: Condensed Matter **19** (2007) 026213.

[9] A. S. Verma and S. R. Bhardwaj, Phys. Stat. Sol. B **243** (2006) 4025.

[10] A. S. Verma, Phys. Stat. Sol. B **246** (2009) 192.

[11] A. S. Verma, Phys. Stat. Sol. B **246** (2009) 345.

[12] A. S. Verma, Philosophical Magazine **89** (2009) 183.

[13] L. Garbato and A. Rucci, Chem. Phys. Letts. **61** (1979) 542.

[14] L. K. Samanta, D. K. Ghosh and G. C. Bhar, Chem. Phys **79** (1983) 361.

[15] D. P. Spitzer, J. Phys. Chem. Solids **31** (1970) 19.

[16] A.V. Ioffe, Soviet Phys. Solid State **5** (1964) 2446.

[17] M. L. Cohen, Phys. Rev. B **32** (1985) 7988.

[18] S.Q. Wang and H. Q. Ye, Phys. Rev. B **66** (2002) 235111.

[19] K. Kim, W. R. L. Lambrecht and B. Segall, Phys. Rev. B **53** (1996) 16310.